\documentclass[sigconf]{acmart}
\usepackage{tikz}
\usepackage{array}     
\usepackage{booktabs}  
\usepackage{array}
\usepackage{tcolorbox}
\usepackage{pifont} 
\usepackage{amsmath}
\usepackage{graphicx}
\usepackage{multirow} 
\usepackage{booktabs}
\usepackage{hyperref}

\usepackage{algorithm}
\usepackage{algpseudocode}
\usepackage{enumerate}
\AtBeginDocument{%
  }
\newcommand{\tstart}{t_{\text{start}}}
\newcommand{\tend}{t_{\text{end}}}
\setcopyright{acmlicensed}
\copyrightyear{2024}
\acmYear{2024}
\acmDOI{XXXXXXX.XXXXXXX}
\acmConference[WWW]{The Web Conference 2025}{April 28--May 02,
  2025}{Sydney, Australia}
\acmISBN{978-1-4503-XXXX-X/18/06}

\begin{document}


\title[TimelineKGQA Generator]{TimelineKGQA: A Comprehensive Question-Answer Pair Generator for Temporal Knowledge Graphs}

\author{Qiang Sun}
\email{pascal.sun@research.uwa.edu.au}
\orcid{0000-0002-4445-0025}
\affiliation{%
  \institution{The University of Western Australia}
  \city{Perth}
  \state{WA}
  \country{Australia}
}

\author{Sirui Li}
\email{sirui.li@murdoch.edu.au}
\orcid{0000-0002-2504-3790}
\affiliation{%
  \institution{Murdoch University}
  \city{Perth}
  \state{WA}
  \country{Australia}
}

\author{Du Huynh}
\email{du.huynh@uwa.edu.au}
\orcid{0000-0003-3080-9655}
\affiliation{%
 \institution{The University of Western Australia}
  \city{Perth}
  \state{WA}
  \country{Australia}
}

\author{Mark Reynolds}
\email{mark.reynolds@uwa.edu.au}
\orcid{0000-0002-5415-0544}
\affiliation{%
 \institution{The University of Western Australia}
  \city{Perth}
  \state{WA}
  \country{Australia}
}

\author{Wei Liu}
\email{wei.liu@uwa.edu.au}
\orcid{0000-0002-7409-0948}
\affiliation{%
 \institution{The University of Western Australia}
  \city{Perth}
  \state{WA}
  \country{Australia}
}

\renewcommand{\shortauthors}{Q Sun et al.}

\begin{abstract}
Question answering over temporal knowledge graphs (TKGs) is crucial for understanding evolving facts and relationships, yet its development is hindered by limited datasets and difficulties in generating custom QA pairs. We propose a novel categorization framework based on timeline-context relationships, along with \textbf{TimelineKGQA}, a universal temporal QA generator applicable to any TKGs. The code is available at: \url{https://github.com/PascalSun/TimelineKGQA} as an open source Python package.
\end{abstract}

\ccsdesc[500]{Do Not Use This Code~Generate the Correct Terms for Your Paper}
\ccsdesc[300]{Do Not Use This Code~Generate the Correct Terms for Your Paper}
\ccsdesc{Do Not Use This Code~Generate the Correct Terms for Your Paper}
\ccsdesc[100]{Do Not Use This Code~Generate the Correct Terms for Your Paper}

\keywords{knowledge graph, temporal knowledge graph, question answering}


\maketitle

\addtolength{\abovedisplayskip}{-1ex}
\addtolength{\belowdisplayskip}{-1ex}

\section{Introduction}

Temporal knowledge graphs~(TKGs) extend standard knowledge graph triples $(e_1, r, e_2)$ to quintuples $(e_1, r, e_2, \tstart, \tend)$ by incorporating temporal stamps. 
TKGQA, temporal knowledge graph question answering, can be seen as a transitional task from knowledge graph information retrieval~(IR) to knowledge graph reasoning, in particular, temporal reasoning.

Firstly, current research in TKGQA is significantly constrained by dataset limitations. 
While CronQuestion~\cite{saxena_question_2021} stands as the largest and most complex TKGQA dataset available, current TKG embedding-based QA methods have already achieved Hits@1 metrics well above 0.9.
More concerning is the trend observed in recent publications: the work proposed by \citeauthor{ijcai2023p571}~\cite{ijcai2023p571} in 2023 achieved 0.969 for Hits@1, followed by two \textbf{suspicious} works by \citeauthor{MT3TQA}~\cite{MT3TQA} and \citeauthor{Xue_Liang_Wang_Zhang_2024}~\cite{Xue_Liang_Wang_Zhang_2024} in 2024 that present nearly identical methodological pipelines and benchmark results without cross-referencing. 
Notably, the ablation study diagram in \cite{Xue_Liang_Wang_Zhang_2024} is identical to that of \cite{ijcai2023p571}. 
This situation not only highlights potential academic integrity concerns but also exemplifies how current dataset limitations are effectively blocking advancement in TKGQA research.

Secondly, from the dataset perspective, since the introduction of TempQuestions~\cite{jia_tempquestions_2018}, the TKGQA research community has recognized the need to address complex temporal questions, yet approaches to modeling this complexity remain divergent. 
While TempQuestions initially considered \textit{ordinal} and \textit{implicit} questions as complex, subsequent datasets~\cite{chen_dataset_2021,jia_complex_2021,saxena_question_2021,DBLP:conf/www/JiaCW24} explored different dimensions including \textit{Before/After} relationships, \textit{First/Last} ordering, and \textit{Aggregation} operations. 
However, no existing dataset comprehensively incorporates all these complexity dimensions - even CronQuestion~\cite{saxena_question_2021}, the largest available dataset, lacks temporal aggregation questions such as \textit{How long in total did Obama and Bush served as president of the US?}, as summarized in Table~\ref{tab:public_datasets}.
This suggests the fundamental challenge lies in developing a comprehensive categorization framework that can characterize various dimensions of temporal complexity within a unified dataset.

\begin{table}[t!]
\centering
\caption{Summary of existing TKGQA datasets}
\label{tab:public_datasets}
\vspace{-10pt}
\resizebox{\linewidth}{!}{ 
\begin{tabular}{c|c|c|c|c|c}
\toprule
\textbf{} & \textbf{TempQuestions} & \textbf{Time-Sensitive QA} & \textbf{TimeQuestions} & \textbf{CronQuestions} & \textbf{TIQ} \\
\midrule
Author & \citet{jia_tempquestions_2018} & \citet{chen_dataset_2021} & \citet{jia_complex_2021} & \citet{saxena_question_2021} & \citet{DBLP:conf/www/JiaCW24a}  \\
Year & 2018 & 2021 & 2021 & 2021 & 2024 \\
Size & 1,364 & 20,000 & 16,181 & 410k & 10,000 \\
Source & 3 KGQA datasets & WikiData & 8 KGQA datasets & WikiData & WikiData \\
\midrule
Explicit/Implicit & \ding{51} & \ding{51} & \ding{51} & \ding{55} & \ding{51} \\
Simple Time & \ding{51} & \ding{51} & \ding{51} & \ding{51} & \ding{51} \\
Simple Entity & \ding{55} & \ding{55} & \ding{55} & \ding{51} & \ding{55} \\
Before/After & \ding{55} & \ding{55} & \ding{55} & \ding{51} & \ding{55} \\
Time Join & \ding{55} & \ding{55} & \ding{55} & \ding{51} & \ding{55} \\
First/Last & \ding{55} & \ding{55} & \ding{55} & \ding{51} & \ding{55} \\
Ordinal & \ding{51} & \ding{51} & \ding{51} & \ding{55} & \ding{55} \\
Aggregation & \ding{55} & \ding{51} & \ding{55} & \ding{55} & \ding{55} \\
\bottomrule
\end{tabular}
}
\vspace{-10pt}
\end{table}

\begin{figure*}[t!]
    \centering
    \includegraphics[width=0.9\linewidth]{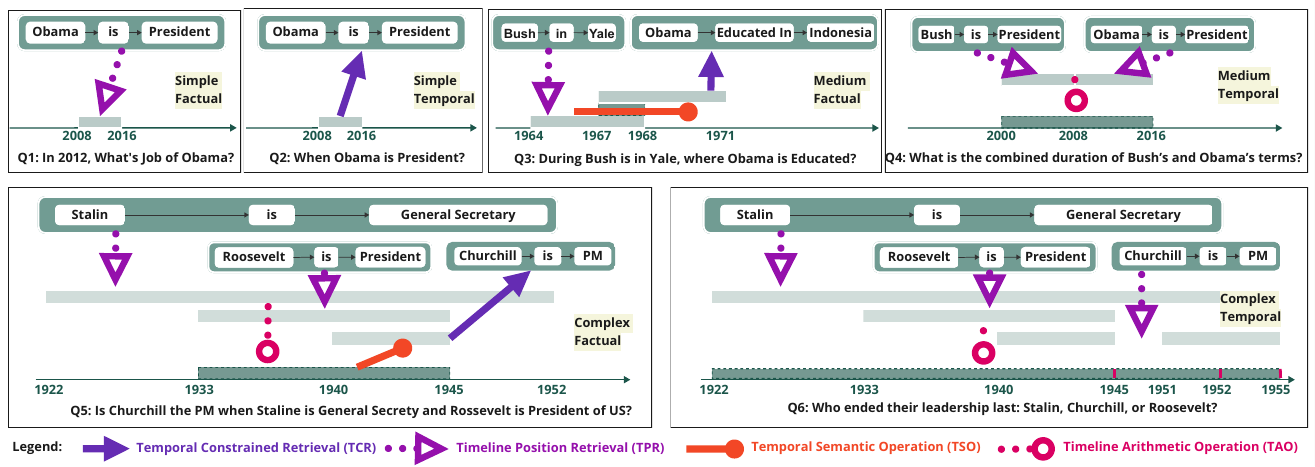}
    \caption{Categorization framework for Temporal Questions}
    \label{fig:timelineqa}
\end{figure*}

Thirdly, while Large Language Models offer promising directions for domain-specific TKGQA applications through fine-tuning, the lack of effective temporal QA pair generation methods for private domains has left this potential largely unexplored.

We propose a comprehensive temporal question categorization framework by treating timeline linearly and homogeneously, addressing complexity through multiple dimensions: (1)~context complexity (Simple, Medium, Complex), (2)~answer focus (Temporal vs. Factual), (3)~temporal relations (Thirteen Allen Temporal Relations, Three Time Range Set Relations, Duration Comparisons and Ranking), and (4)~temporal capabilities, grouped into temporal retrieval capabilities (temporal constrained retrieval~(TCR) and timeline position retrieval~(TPR)) and temporal operation capabilities (temporal semantic operation~(TSO) and timeline arithmetic operation~(TAO)).
Based on this framework, we develop a Python package that can generate TKGQA datasets incorporating all these complexity dimensions from any given TKG, supporting answer formats including not only traditional entity and timestamp answers, but also Yes/No, time range, and duration answers.

\section{Related Work}
TempQuestions~\cite{jia_tempquestions_2018}, one of the early TKGQA datasets, introduced basic categories including \textit{Explicit Temporal}, \textit{Implicit Temporal}, \textit{Temporal Answer}, and \textit{Ordinal Constraints}. 
CronQuestions~\cite{saxena_question_2021} further developed this categorization with \textit{Simple Time}, \textit{Simple Entity}, \textit{Before/After}, \textit{First/Last}, and \textit{Time Join}, essentially extending temporal relations through a fuzzy version of Allen's temporal logic. 
Complex-CronQuestions~\cite{jia_complex_2021} focused on complex questions, while the recent \textit{Temporal Implicit Questions}~(TIQ)~\cite{DBLP:conf/www/JiaCW24a} dataset emphasized challenging implicit questions.
However, as shown in Table~\ref{tab:public_datasets}, these datasets not only lack comprehensive coverage of temporal question types (e.g., \textit{Aggregation}, \textit{Ordinal}, \textit{First/Last}), but more fundamentally, they fail to establish a systematic framework for characterizing temporal complexity across different dimensions. 
This absence of a multi-dimensional temporal complexity framework, combined with incomplete coverage of temporal operations, significantly limits our ability to advance TKGQA research.

\begin{figure*}[t!]
    \centering
    \includegraphics[width=0.9\textwidth]{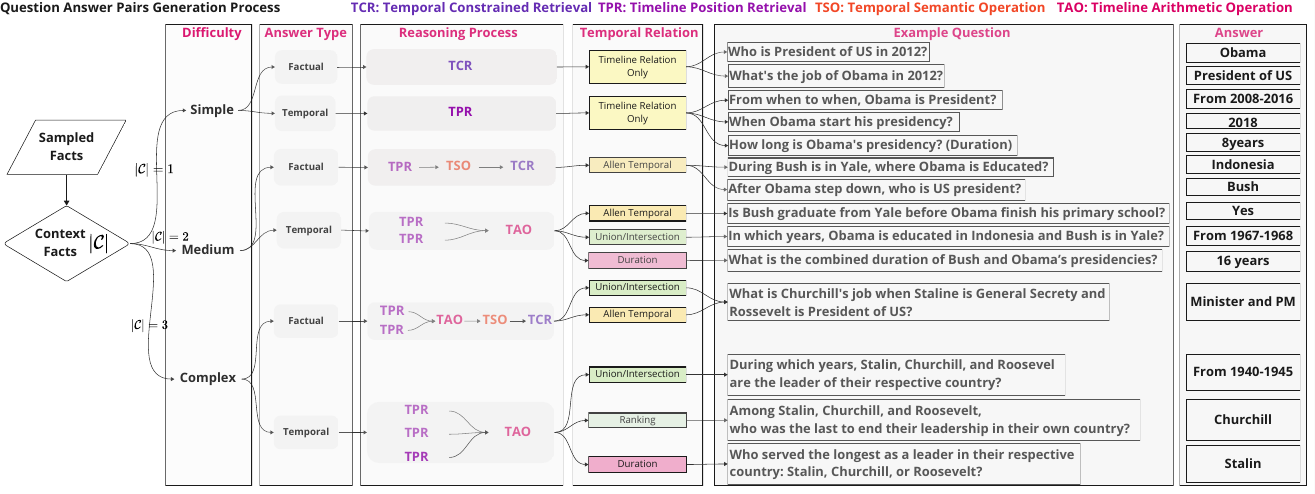} \\[-2ex]
    \caption{Question answer pair generation procedure in our TimelineKGQA Generator}
    \label{fig:qa-generator}
    \vspace{-2ex}
\end{figure*}

\section{Concepts and notation}
\textbf{TKG. }A TKG, denoted as $\mathcal{G} := (\mathcal{E}, \mathcal{R}, \mathcal{T}, \mathcal{F})$, is a directed graph that incorporates temporal relationships between entities. In this notation, $\mathcal{E}$ represents the set of entities, $\mathcal{R}$ denotes the set of relations, $\mathcal{T}$ signifies the set of timestamps, and $\mathcal{F}$ is the set of facts. Each fact is represented as $f = (e_{1}, r, e_{2}, \tstart, \tend)$, where $f \in \mathcal{F}$; $e_{1}, e_{2} \in \mathcal{E}$; $r \in \mathcal{R}$ and $\tstart, \tend \in \mathcal{T}$.

\noindent
\textbf{TKGQA and Context Facts. }Given a TKG $\mathcal{G}$ and a natural language question $q$, TKGQA aims to find the relevant facts in $\mathcal{F}$ that can answer $q$. These relevant facts are referred to as \textit{context facts}, represented as a set $\mathcal{C}=\{f_1, f_2, ...\} \subset \mathcal{F}$ where $|\mathcal{C}|$ is the number of contextual facts for $q$.

\section{Question Complexity Dimensions}
Our categorization framework classifies question-answer pairs based on four dimensions: context complexity ($|\mathcal{C}|$), answer focus, temporal relations, and temporal capabilities, with detailed illustrations shown in Figures~\ref{fig:timelineqa} and~\ref{fig:qa-generator}.

\vspace{0.5em}
\noindent\textbf{Simple (if $|\mathcal{C}|=1$):}
Questions requiring a single context fact are further categorized by their answer focus. \textit{Factual} questions need temporal constrained retrieval~(TCR) to find facts within given time constraints \((t'_{\text{start}}, t'_{\text{end}})\), while \textit{Temporal} questions require timeline position retrieval~(TPR) to determine time ranges \((t_{\text{start}}, t_{\text{end}})\) of given facts. In Figure~\ref{fig:timelineqa}, Q1 exemplifies \textit{Simple.Factual} and Q2 represents \textit{Simple.Temporal}.
\vspace{0.5em}

\noindent\textbf{Medium (if $|\mathcal{C}|=2$):}
Questions involving two context facts require more complex temporal capabilities. For \textit{Factual} questions, both TPR and TCR are needed, along with temporal semantic operation~(TSO). For example, Q3 in Figure~\ref{fig:timelineqa} first uses TPR to retrieve Bush's Yale period \(T = (1964,1968)\), then applies TSO with the temporal signal "during" to infer a new time range $T' = (1964,1968)$, and finally uses TCR with the new $T'$ to retrieve Obama's education facts within this period.
TSO generates a new time range $T'$ based on a time range \(T = (\tstart, \tend)\) and a temporal signal word $w$ (e.g., \textit{before}, \textit{after}, \textit{during}):
\begin{equation}
T' = \textit{op} (w, \tstart, \tend)
\label{eq:op}
\end{equation}

For \textit{Temporal} questions like Q4, TPR will be first applied to retrieve relevant time ranges (e.g., presidency periods of Bush and Obama), then \textit{Timeline Arithmetic Operation}~(TAO) processes these ranges through three types of temporal relation operations: (1)~\textbf{Set Operations} (\textit{union}, \textit{intersection}, \textit{negation}), (2)~\textbf{Allen's Temporal Relations} (13 relations like \textit{before}, \textit{after}, \textit{start-by}), and (3)~\textbf{Duration Operations} (duration comparison and calculation). For Q4, applying the 'union' operation to the presidency periods yields a 16-year duration.

\vspace{0.5em}
\noindent\textbf{Complex (if $|\mathcal{C}|=3$):} 
Under this category, \textit{Factual} questions require all four capabilities (TPR, TCR, TSO, and TAO) as shown in Q5 in Figure~\ref{fig:timelineqa}, while \textit{Temporal} questions need only TPR and TAO (Q6). The key distinction from the \textbf{Medium} category is the introduction of \textit{Ordinal} type questions, which require a fourth type of temporal relation operation in TAO: (4)~\textbf{Ranking} (temporal ordering based on $\tstart$, $\tend$, or \textit{duration}).

\begin{table}[t!]
    \centering
    \caption{Recategorize CronQuestions using our framework}
    \resizebox{0.95\linewidth}{!}{
    \begin{tabular}{lrrlr}
        \toprule
        \multicolumn{3}{c}{\textbf{Difficulty and Counts}} & \multicolumn{2}{c}{\textbf{Question categorisation}} \\
        \cmidrule(r){1-3} \cmidrule(l){4-5}
        \textbf{Difficulty} & \textbf{Template Count} & \textbf{Question Count} & \textbf{Category} & \textbf{Count} \\
        \midrule
        \multirow{2}{*}{Simple} & \multirow{2}{*}{158} & \multirow{2}{*}{177,922} & Simple.Factual & 106,208 \\
        & & & Simple.Temporal & 71,714 \\
        \midrule
        \multirow{2}{*}{Medium} & \multirow{2}{*}{165} & \multirow{2}{*}{90,641} & Medium.Factual & 90,641 \\
        && & \textbf{\color{red}{Medium.Temporal}} & \textbf{\color{red}{0}} \\
        \midrule
        \multirow{2}{*}{Complex} & \multirow{2}{*}{331} & \multirow{2}{*}{141,437}  & Complex.Factual & 67,709 \\
        && & Complex.Temporal & 73,728 \\
        \midrule
        Total & 654 & 410,000 & & 410,000 \\
        \bottomrule
    \end{tabular}
    }
    \label{tab:cronkg-class}
\end{table}

After reclassifying CronQuestion with our proposed categorization framework (Table~\ref{tab:cronkg-class}), several limitations emerge. While the dataset contains numerous templates (654), it is dominated by \textbf{Simple} questions with structurally similar patterns. More importantly, it lacks diversity in temporal relation complexity: the \textit{Medium.Temporal} category is entirely absent, ranking questions are limited to "first" and "last" operations, and Allen's temporal relations only cover basic relationships (\textit{before}, \textit{after}, \textit{during}). The dataset also lacks duration comparisons and time range inference questions.

\section{Question Answer Pair Generator}
Based on the aforementioned categorization framework, we develop \textbf{TimelineKGQA}, an open-source automated QA pair generator for temporal knowledge graphs. 
This tool can generate diverse QA pairs from any TKG while incorporating all identified complexity dimensions. 
As shown in Figure~\ref{fig:qa-generator}, our \textbf{TimelineKGQA} consists of four main modules:

\begin{figure}[t!]
    \centering
    \includegraphics[width=0.85\linewidth]{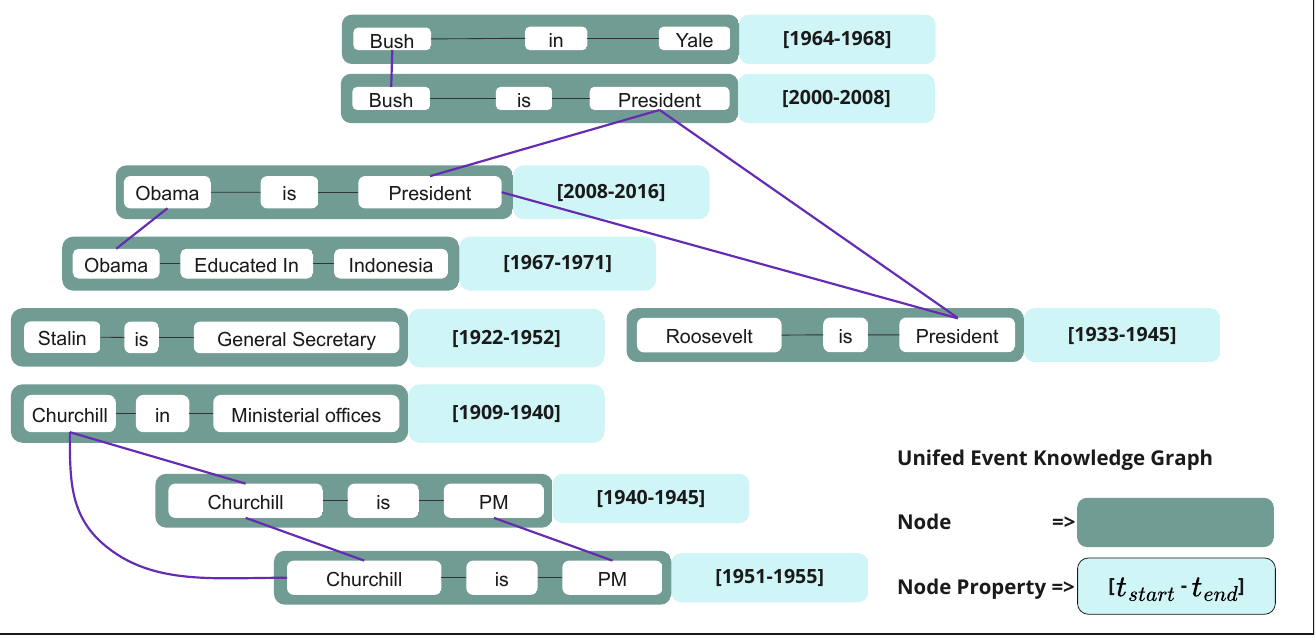} \\[-2ex]
    \caption{Temporal knowledge graph unification}
    \label{fig:qa-kg-transformation}
    \vspace{-2ex}
\end{figure}

\noindent\textbf{Module 1: Temporal knowledge graph unification.}
Our generator accepts any knowledge graph as input, converting it into a TKG where each fact is augmented with a time range $(\tstart, \tend)$ (Figure~\ref{fig:qa-kg-transformation}). Unlike CronQuestions which only supports year-level granularity, our framework allows for flexible temporal granularity (minutes, hours, days to years) and supports both time ranges and instant events (where $\tstart = \tend$, can be treated as a time point).

\noindent\textbf{Module 2: Fact Sampling.}
To mimic real-world question patterns, our sampling strategy for context facts ($|\mathcal{C}| \in \{1, 2, 3\}$) prioritizes temporally proximate facts and frequently occurring entities.

\noindent\textbf{Module 3: Question Answer Pair Generation.}
Based on the number of sampled facts, the generator follows different generation paths (Figure~\ref{fig:qa-generator}):

\noindent\textbf{Simple} ($|\mathcal{C}|=1$): Five questions are generated for each fact - two factual focus questions about entities, and three temporal focus questions involving $\tstart$, $\tend$, or the duration.

\noindent\textbf{Medium} ($|\mathcal{C}|=2$): For \textit{Medium.Factual} answers, Allen's Temporal Relations between two facts are calculated and incorporated as temporal signal words in the questions, with answers about entities or predicates. For \textit{Medium.Temporal} answers, questions are generated based on one of three temporal relation operations (union/intersection, Allen's relations, or duration), using Yes/No, multiple-choice or open-end answer format.

\noindent\textbf{Complex} ($|\mathcal{C}|=3$): Follows similar generation process, with additional support for ranking-based questions where fact rankings are calculated before question formulation.

\noindent\textbf{Module 4: LLM Paraphrase.} 
To avoid the limitations of template-based questions (e.g., unnatural expressions and trivial entity matching seen in CronQuestions), we use an LLM to paraphrase questions into more natural and diverse expressions. The code for \textbf{TimelineKGQA} is available via GitHub\footnote{\url{https://github.com/PascalSun/TimelineKGQA}}.

\section{Dataset Generation}
Using TimelineKGQA, we create two benchmark datasets from the ICEWS Coded Event Data\footnote{\url{https://dataverse.harvard.edu/dataset.xhtml?persistentId=doi:10.7910/DVN/28075}}~(Time Range) and CronQuestion knowledge graph~(Time Point) for demonstration purposes, generating 89,372 and 41,720 questions respectively. The distribution of questions across categories and answer types for both \textit{ICEWS Actor} and \textit{CronQuestion KG} datasets is shown in Tables~\ref{tab:combined_distribution} and~\ref{tab:combined_answer_types}.

\begin{table}[hbt]
    \centering
    \caption{Questions distribution across Train/Val/Test sets and required temporal capabilities for \textit{ICEWS Actor} and \textit{CronQuestion KG}.}
    \label{tab:q-distrib}
   \vspace{-2ex}    
    \resizebox{\linewidth}{!}{
    \begin{tabular}{llrrr|lr}
        \toprule
        \textbf{Source KG} & & \textbf{Train} & \textbf{Val} & \textbf{Test} & \textbf{Temporal capabilities} & \textbf{Count} \\
        \midrule
        \multirow{4}{*}{\textit{ICEWS Actor}} & Simple & 17,982 & 5,994 & 5,994 & Temporal Constrained Retrieval & 34,498 \\
        & Medium & 15,990 & 5,330 & 5,330 & Timeline Position Retrieval & 79,382 \\
        & Complex & 19,652 & 6,550 & 6,550 & Timeline Operation & 34,894 \\
        &  &  &  &  & Temporal Semantic Operation & 24,508 \\
        \cmidrule{2-7}
        & \textbf{Total} & \textbf{53,624} & \textbf{17,874} & \textbf{17,874} &  & 89,372 \\
        \midrule
        \multirow{4}{*}{\textit{CronQuestion KG}} & Simple & 7,200 & 2,400 & 2,400 & Temporal Constrained Retrieval & 19,720 \\
        & Medium & 8,252 & 2751 & 2751 & Timeline Position Retrieval & 37,720 \\
        & Complex & 9580 & 3,193 & 3,193 & Timeline Arithmetic Operation & 21,966 \\
        &  &  &  &  & Temporal Semantic Operation & 15,720 \\
        \cmidrule{2-7}
        & \textbf{Total} & \textbf{25,032} & \textbf{8,344} & \textbf{8,344} &  & 41,720 \\
        \bottomrule
    \end{tabular}
    }
    \label{tab:combined_distribution}
\end{table}

\begin{table}[hbt]
    \centering
    \caption{Distribution of various detailed answer types for ICEWS Actor and CronQuestions KG}
    \label{tab:combined_answer_types}
    \vspace{-2ex}    
    \resizebox{0.8\linewidth}{!}{
    \begin{tabular}{l rr}
        \toprule
        \textbf{Detailed answer types} & \textbf{ICEWS Actor} & \textbf{CronQuestions KG} \\
        \midrule
        Subject & 17,249 & 9,860 \\
        Object & 17,249 & 9,860 \\
        Timestamp Start & 4,995 & 2,000 \\
        Timestamp End & 4,995 & 2,000 \\
        Timestamp Range & 4,995 & 2,000 \\
        Duration & 4,995 & 2,000 \\
        Relation Duration & 9,971 & 4,000 \\
        Relation Ranking & 4,981 & 2,000 \\
        Relation Union or Intersection & 19,942 & 8,000 \\
        \bottomrule
    \end{tabular}
    }
   \vspace{-2ex}        
\end{table}

\vspace{0.5em}
Due to the unique characteristics of our more challenging datasets, particularly the presence of time range inputs and diverse answer formats beyond simple entity/timestamp pairs, existing embedding-based TKGQA models (which typically only handle time points as input and output either timestamps or entities) cannot be directly applied. 
To evaluate whether question difficulty aligns with our complexity categorization, we implement a Retrieval Augmented Generation~(RAG) baseline. 
Facts are first encoded as natural language statements following the format 
\begin{tcolorbox}[boxrule=0pt,colback=gray!20,left=2pt,right=2pt,top=2pt,bottom=2pt]
\textit{\{subject\} \{predicate\} \{object\} from \{$\tstart$\} to \{$\tend$\}}
\end{tcolorbox}
\noindent and embedded using OpenAI's \textit{text-embedding-3-small} model. 
Queries are embedded similarly, followed with semantic similarity search retrieving the top \( K \) relevant facts. 
We evaluate the retrieval performance using modified versions of Mean Reciprocal Rank~(MRR) and Hits@K metrics, where for questions requiring multiple context facts (2 facts for \textit{Medium}, 3 for \textit{Complex}), all context facts must be retrieved within the top \(|\mathcal{C}| \cdot K\) positions.

\begin{table}[t]
   \centering
   \caption{RAG baseline performance. O: Overall, S: Simple, M: Medium, C: Complex.}
    \vspace{-2ex}  
   \label{tab:rag-results}
   \setlength{\tabcolsep}{4pt}
      \resizebox{0.8\linewidth}{!}{
   \begin{tabular}{|l|cccc|cccc|}
       \hline
       & \multicolumn{4}{c|}{ICEWS Actor} & \multicolumn{4}{c|}{CronQuestions KG} \\
       \cline{2-9}
       & O & S & M & C & O & S & M & C \\
       \hline
       MRR $\uparrow$ & 0.365 & 0.726 & 0.274 & 0.106 & 0.331 & 0.771 & 0.218 & 0.101 \\
       Hits@1 $\uparrow$ & 0.265 & 0.660 & 0.128 & 0.011 & 0.235 & 0.704 & 0.092 & 0.009 \\
       Hits@3 $\uparrow$ & 0.391 & 0.776 & 0.331 & 0.086 & 0.348 & 0.824 & 0.249 & 0.077 \\
       \hline
   \end{tabular}
   }
   \vspace{-3ex}  
\end{table}
The results are reported in Table~\ref{tab:rag-results}, showing that our categorisation framework effectively reflects question difficulty, as RAG's performance consistently decreases from \textbf{Simple} to \textbf{Medium} and \textbf{Complex} categories (e.g., Hits@1 on ICEWS Actor drops from 0.660 to 0.128 and 0.011 respectively).

\section{Conclusion}
We present a comprehensive temporal question categorization framework and a universal TKGQA dataset generator that addresses the limitations of existing datasets. 
Our framework introduces a systematic approach to classify question complexity through context facts, answer focus, and temporal operations, while identifying four key temporal capabilities (TCR, TPR, TSO, and TAO). 
Using our \textbf{TimelineKGQA} generator to create two benchmark datasets, we demonstrate that question difficulty aligns with our complexity categorization through empirical evaluation. 
This work provides a foundation for developing and evaluating more advanced TKGQA solutions, and enables widespread application in private domains.

\bibliographystyle{ACM-Reference-Format}
\bibliography{ref}

\end{document}